\documentclass{article}

\usepackage{arxiv}

\usepackage[utf8]{inputenc} 
\usepackage[T1]{fontenc}    
\usepackage{hyperref}       
\usepackage{url}            
\usepackage{booktabs}       
\usepackage{amsfonts}       
\usepackage{nicefrac}       
\usepackage{microtype}      
\usepackage{lipsum}
\usepackage{graphicx}
\usepackage{natbib}
\graphicspath{ {./images/} }

\title{On the feasibility of dynamical analysis of network models of biochemical regulation}

\author{
Luis M. Rocha,$^{1,2,3}$ \\
$^{1}$Department of Systems Science and Industrial Engineering,\\ Binghamton University, State University of New York, Binghamton, NY, USA.\\
$^{2}$Instituto Gulbenkian de Ciência, 2780-156, Oeiras, Portugal \\
$^{3}$Center for Social and Biomedical Complexity, \\
Luddy School of Informatics, Computing \& Engineering, \\
Indiana University, Bloomington IN, USA\\
  \texttt{rocha@binghamton.edu}
}

\begin{document}
\maketitle



To the Editor: A recent article by \cite{weidner2021capturing} presents a method to extract graph properties that are predictive of the dynamical behavior of multivariate, discrete models of biochemical regulation. In other words, a method that uses only features from the structure of network interactions to predict which nodes are most involved in automata network dynamics.
However, the authors claim that dynamical analysis of large automata network models is ``not even feasible.''
To make sure that others are not discouraged from working on this problem, it is important to clarify that effective dynamical analysis of automata network models, to the contrary, is feasible. Unlike what is suggested in the article, graph-based analysis of static features is not the only analytical avenue for large systems biology models of regulation and signaling dynamics because there are dynamical methods that are, indeed,
%
\textit{scalable}\footnote{
%
%
By scalable we mean that the computational complexity of methods employed to analyze multivariate dynamical systems in regard to their dynamical behavior (e.g., controlability, convergence to attractors, robustness to perturbations, etc.) is manageable. That is, a given method is scalable if results can be computed in finite (and reasonable) time, with finite memory, for a system of a reasonable size---ideally for network models in Systems Biology, up to thousands of nodes.
%
}.

There has been much interest recently in predicting multivariate dynamics from static network structure alone, especially in regard to the controlability of systems biology models of gene regulation, signalling, and cellular differentiation \citep{Liu2011control, nacher2013structural,Fiedler2013fvc, Zanudo2017fvc}.
These are quite welcome methods because we often lack information about the underlying causal interaction dynamics. However,
%
two very popular (and scalable) methods in network science
%
lead to very erroneous predictions of the subsets of (driver) variables that control dynamics \citep{Gates2016bncontrol}.
%
The most accurate of these methods are based on feedback vertex set theory  \citep{Fiedler2013fvc,Zanudo2017fvc}, which does not scale well and
%
can only make predictions about the entire ensemble of dynamical systems that fits the same static interaction graph \citep{gates2021effective}.

Another putative reason for pursuing structure-only methods is that even when the underlying interaction dynamics of each variable is known, it is not feasible to compute the dynamical (or attractor) landscape of the entire multivariate system when the interaction network is sufficiently large. This prevents us from exhaustively  enumerating all possible interventions that can control the dynamics from one attractor basin to another\footnote{
%
The most important scalability constraint in the analysis of automata networks is the number of node variables, $n$, as the dynamical landscape of such systems is comprised of $s^n$ possible configurations of states $s$ \citep{Gates2016bncontrol}. For instance, a well-known Boolean network model of intracellular signaling networks in generic fibroblasts is comprised of 130 nodes \citep{helikar_emergent_2008}, thus it can be in one of $2^{130}$ possible state configurations---a dynamical landscape that is too large to be exhaustively searched.
}.
Still, it is also true that enumeration of all possible interventions is infeasible in a simple graph of sufficient size. The computational complexity of finding all possible subsets of the set of nodes, or generating the powerset, is at least $\mathcal{O}(2^n)$ \citep{Moore_bounds}. Therefore, exhaustive search of all possible interventions in any network is ultimately infeasible for large graphs whether one uses structure- or dynamics-based methods.

Computing the full dynamics of a multivariate system certainly adds to the complexity of exhaustively searching all possible interventions.
%
But many software tools exist---and are collected in repositories such as the \texttt{CoLoMoTo} Consortium \citep{naldi2015cooperative}---that allow for the identification of attractors in automata networks without full enumeration of their dynamical landscapes, such as \texttt{PyBoolNet} \citep{PyBoolNet} and \texttt{Boolink} \citep{karanam2021boolink}.
In particular, recent developments in attractor identification algorithms and code (\texttt{PyStableMotifs}) now allow the dynamical analysis of thousands of networks, some with over 15,000 nodes \citep{rozum2021parity}.
From another angle, a novel update scheme for asynchronous automata networks has been shown to reduce the complexity of attractor identification, enabling the modelling and dynamical analysis of genome-scale networks \citep{pauleve2020reconciling}.

Moreover, various computational approaches have been developed and applied to extract the key drivers of collective dynamics of biochemical network models
%
without going through every possible subset of nodes, much less the entire dynamical landscape, in a brute force manner  \citep{Harie2103698118,Zanudo2015cellfate, biane2018causal, su2020dynamics, rozum2021parity}.
%
Indeed, scalable methods exist that remove the redundancy of the dynamics of each variable (micro-level) to allow for a characterization of the entire causal macro-level dynamics, in both complete \citep{MarquesPita2013canalization} and probabilistic \citep{gates2021effective} manners.
These scalable methods,
%
and the associated software tool \texttt{CANA} \citep{Correia2018cana},
%
provide causal graph representations of automata networks that synthesize both structure and dynamics. They are exhaustive in the sense that they preserve all effective interactions of the micro-level dynamics---only redundant interactions are disregarded. Thus, it is very feasible to analyze systems biology models without disregarding their dynamics, allowing the precise study of any putative intervention that controls the dynamics just as easily as structure-only methods, but with additional accuracy afforded by information about the dynamics.

\section*{Acknowledgements}
The author is indebted to the anonymous reviewers for their most valuable comments and references which have considerably strengthened this letter.

\bibliographystyle{abbrvnat}
\bibliography{document}


\end{document}